\documentclass[doublecol]{epl2} 
\usepackage{amssymb}
\usepackage{amsmath}

\newcommand{\beq}{\begin{equation}}
\newcommand{\eeq}{\end{equation}}
\newcommand{\bea}{\begin{eqnarray}}
\newcommand{\eea}{\end{eqnarray}}

\newcommand{\al}{\alpha}

\newcommand{\iom}{i\omega_n}

\newcommand{\be}{\beta}

\title{State-of-the-art techniques for calculating spectral functions in models for correlated materials}
\shorttitle{Title} 

\author{K. Hallberg, D. J. Garc\'{\i}a, Pablo S. Cornaglia,  Jorge I. Facio and Y. N\'u\~nez-Fern\'andez }
\institute{                    
	{Centro At{\'o}mico Bariloche and Instituto Balseiro, CNEA and CONICET, 8400 Bariloche, Argentina}\\

}

\shortauthor{Y. Nu\~nez Fern\'andez \etal}

\pacs{71.27.+a}{Strongly correlated electron systems; heavy fermions}
\pacs{75.40.Mg}{Numerical simulation studies}
\pacs{71.15.-m}{Methods of electronic structure calculations}

\abstract{
The dynamical mean field theory (DMFT) has become a standard technique for the study of strongly correlated models and materials overcoming some of the limitations of density functional approaches based on local approximations. An important step in this method involves the calculation of response functions of a multiorbital impurity problem which is related to the original model. Recently there has been considerable progress in the development of techniques based on the density matrix renormalization group (DMRG) and related matrix product states (MPS) implying a substantial improvement to previous methods. 
In this article we review some of the standard algorithms and compare them to the newly developed techniques, showing examples for the particular case of the half-filled two-band Hubbard model.
}

\begin{document}

\maketitle
\section{Introduction}
The research on materials having strong electron-electron correlations due to interactions in local orbitals has attracted a great deal of attention in recent years. This is due to their fascinating properties like high temperature superconductivity, colossal magnetoresistance or heavy fermion behavior, and their sensitivity to external fields which makes them attractive in view of applications. 
In these materials, strong electron correlations play a central role and represent a major challenge for the understanding and control of the different phenomena. In spite of the important success of the methods based on Density Functional Theory (DFT) \cite{hohenberg} in the electronic structure calculations of weakly correlated materials, major difficulties are found when dealing with {\it f} or {\it d} electron systems where the screened local interaction energy is of the order of the conduction electron bandwidth. The DFT-based local density approximation (LDA) \cite{jones} and its generalizations are unable to describe accurately the strong electron correlations.

To overcome these difficulties, the Dynamical Mean-Field Theory (DMFT) and its cluster versions were developed \cite{pt,review,savrasov2001,maier,hettler,senechal}, which allow to extend these methods and treat the dynamical electron correlations in a reliable way. The DMFT has become one of the basic methods to calculate realistic electronic band structure in strongly correlated
systems. The combination of the DMFT with LDA had allowed for band structure calculations of a large variety of correlated materials (for reviews see Refs. \cite{imada,Held}), where the DMFT accounts more reliably for the local correlations \cite{anisimov,katsnelson}.

A recent alternative proposal, the Density Matrix Embedding Theory, DMET, was developed, which relies on the embedding of the wave functions of a local cluster fragment (instead of the local Green functions) in a self-consistent finite environment \cite{Chan1,Bulik} and which seems to be a good alternative to the DMFT. 

A key point of the DMFT method is the solution of an associated quantum impurity problem where the fermionic environment of the impurity has to be determined self-consistently until convergence of the local Green function and the local self-energy is reached. This approach is exact for the infinitely coordinated system (infinite dimensions), the non-interacting model and in the atomic limit. 
Therefore the possibility to obtain
reliable DMFT solutions of lattice Hamiltonians relies directly on the ability to solve (complex) quantum impurity models. 

During the early stages of the development of the DMFT, several quantum impurity solvers were proposed and used successfully  such as the iterated perturbation theory (IPT) \cite{GeorgesKotliar,GeorgesKrauth,ZhangRozenberg,RozenbergKotliar}, exact diagonalization (ED) \cite{Caffarel}, the Hirsch-Fye quantum Monte Carlo (HFQMC) \cite{ZhangRozenberg,hf,mr,Jarrell,Georges2}, non-crossing approximations (NCA) \cite{NCA}, and the numerical renormalization group (NRG)
 \cite{wrg,bulla,Cornaglia2007,vollhardt,vondelft}. These methods were good enough to allow for the calculation of the metal-insulator transitions and other low-lying energy properties. However, they suffered from important drawbacks that had to be overcome if other interesting properties were to be calculated, such as systems having a larger number of bands, other kind of interactions (such as spin-orbit and electron-phonon coupling, etc), and interband hybridization. Among the drawbacks, one can mention the sign problem and the difficulty in reaching low temperatures in the HFQMC algorithm \cite{Loh}, the failure of the NCA in obtaining a reliable solution for the metallic state, the limitation to few lattice sites, far from the thermodynamic limit of the ED and the reduced high-energy resolution of the NRG technique (see improvements to this in  Ref. \cite{Zitko}).

More recently, several other impurity solvers have been developed that overcome (at least partially) many of these problems among which we can mention the Density Matrix Renormalization Group (DMRG) \cite{garciadmft,garciadmft2,yuriel,nishimoto,karski}, the continuous time quantum Monte Carlo (CTQMC) \cite{ctqmc1,ctqmc2,ctqmc3,CTQMC,ryo} and the fluctuation exchange approximation (FLEX) \cite{flex}. 
The CTQMC works at lower temperatures than the HFQMC but still suffers from sign problems for certain models {\it e.g.} with interband hybridization and, most importantly, it also requires an analytical continuation of the Green functions from the imaginary to the real frequency axis which makes it unreliable for some physical quantities involving higher energy bands \cite{Jarrell2}. In addition, FLEX is limited to a certain range of interaction strengths \cite{KotliarSavrasov}.  We will expand on the DMRG variant below.

Other methods proposed as impurity solvers include the equation of motion technique (used in a bath with separate low and high energy degrees of freedom for single and multiple-orbital Hamiltonians) \cite{feng,feng2,litong}, the quasi-continuous-time solver \cite{RostAssad}, an improved IPT approach for large interactions \cite{Tremblay} (which can be compared with the local moment approach, also developed to deal with strong interactions and arbitrary fillings \cite{Barman}), and a two-mode approximation based on Gutzwiller variational approach \cite{Dai}. Also recently proposed are methods based on exact diagonalization (ED) improved by the use of a restricted active basis set for the impurity \cite{Lu}, by a stochastic distribution approach \cite{Granath} and by an augmented version which involves finite temperature and cluster perturbation \cite{Littlewood}. Other promising methods have been proposed based on configuration-interaction approximations to ED and from the quantum chemical perspective \cite{ChanCI}. 

In recent years the so-called slave boson  approach \cite{Barnes1976} in the mean field approximation \cite{Coleman1984,KR1986,li1989spin} has been generalized to preserve the symmetries of the Hamiltonian in the multiorbital case. The rotationally invariant slave-boson mean-field theory (RISB) \cite{Lechermann2007,PhysRevB.80.115120,Ferrero2009a,Ferrero2009b}, provides a real-axis description of the low energy excitations of the system. It's main advantages are the lack of finite size effects and the speed at which solutions of the quantum impurity problem can be found. The lattice version of RISB is equivalent to the multiband Gutzwiller wave-function approximation \cite{Bunemann1998}. Recently, these low energy techniques have been interfaced with LDA calculations \cite{0953-8984-22-27-275601,PSSB:PSSB201147052,PhysRevB.86.045130,PhysRevB.85.035133,PhysRevB.90.161104,PhysRevB.91.125148,PhysRevX.5.011008}. In the sections below, we present the RISB formalism an compare some results with those calculated within the DMRG framework. 

A recent important extension of the DMFT equations concerns its application to treat problems out of equilibrium, extending it to the Keldysh formalism\cite{Monien,freericks,aoki}. In this context, an interesting improvement\cite{gramsch} performs an exact mapping of the action of the equations onto a single impurity Anderson model with time-dependent parameters.   

For newcommers to the field, it is recommendable to visit the TRIQS and ALPS code libraries containing state-of-the-art methods for solving interacting quantum systems \cite{triqs,alps}.

\section{The DMRG as an impurity solver of the DMFT}

It has been shown that the Density Matrix Renormalization Group (DMRG)\cite{white1, book, scholl, karen1} can be used very reliably to solve the related impurity
problem within DMFT\cite{garciadmft, garciadmft2, yuriel}. By using the DMRG to solve the related
impurity problem, the density of states is obtained directly on the real axis (or with a very small imaginary offset) being this its major advantage as compared, for example, to QMC techniques. 
In addition, no \textit{a priori} approximations are made and the
method provides equally reliable solutions for both gapless and gapfull
phases. More significantly, it provides accurate estimates for the
distribution of spectral intensities of high frequency features such as the
Hubbard bands and their structure, which are of main relevance for analysis of x-ray
photoemission and optical conductivity experiments, among others. 

Other techniques using alternative methods for the calculation of dynamical properties within the DMRG have been proposed \cite{nishimoto,karski} and, more recently, methods using the time evolution DMRG algorithm (time evolving block decimation, TEBD) \cite{vidaltebd} for the one- and two-orbital models where shown in \cite{ganahl} (see below). 

In the context of the DMRG and the related matrix product states (MPS) as impurity solvers within the DMFT, recent developments include the Kernel Polynomial Method (Chebyshev expansion for Green functions) \cite{weisse,wolf,ganahl2},  the block Lanczos approach \cite{shirakawa}, a pole decomposition technique within the correction-vector method for the dynamics \cite{peters} and the application to non-equilibrium DMFT using MPS \cite{wolf2}. In this work the authors explore other geometries for the impurity bath showing an increased efficiency for the star environment. 

It was recently realized that converging the DMFT loop on the the imaginary-frequency axis rather than the real-frequency axis reduces computational costs by orders of magnitude because the bath can be represented in a controlled way with far fewer bath sites and, crucially, the imaginary-time evolution does not create entanglement. The imaginary time setup can therefore treat much more sophisticated model Hamiltonians, opening the possibility of studying more complicated and realistic models and performing cluster dynamical mean field calculations for multiorbital situations. The price to be paid is a reduced resolution on the real-frequency axis \cite{Go}.

\section{Model and implementation}
As an example of an implementation of the DMRG impurity solver, in this paper we describe the half-filled two-orbital Hubbard model on a square lattice including a Hund's coupling:
\begin{equation}
H=\sum_{i}h_{i}-t\sum_{<ij> m \sigma}\left(c_{i m \sigma}^{\dagger}c_{j m \sigma}\right)\mbox{,}
\label{eq:ham}
\end{equation}
where $i,j$ are the sites of a square lattice and brackets indicate nearest neighbors, $m$ indicates each of the two orbitals and $\sigma$ is the spin of the electron, whose creation and destruction operators are $c^{\dagger}$ and $c$, respectively.

Defining $n_{i}$ and $s_{i}$ as the on-site charge and spin operators respectively, a rotationally invariant on-site Hamiltonian is:

\begin{equation}\label{eq:hi}
h_{i}=\frac{U}{2}n_{i}^{2}+\frac{J}{2}s_{i}^{2}-{\mu} n_{i}\mbox{.}
\end{equation}

For $\mu=2U$ and ferromagnetic J (J$<$ 0) the ground state of $h_i$ is a triplet and the total Hamiltonian reads: 
\begin{eqnarray}
H&=&\left(U-\frac{3}{4}J\right)\sum_{i m}n_{i m\uparrow}n_{i m\downarrow}+U\sum_{i \sigma \sigma'}n_{1\sigma}n_{2\sigma'} \nonumber \\
&+&J\sum_{i}\mathbf{S}_{i1}\cdot\mathbf{S}_{i2} 
-t\sum_{<ij> m \sigma}\left(c_{i m \sigma}^{\dagger}c_{j m \sigma}\right) \nonumber \\
&+& \left(\frac{-3}{2}U+\frac{3}{8}J\right)\sum_{i} n_{i}
\mbox{,}\label{eq:H-1}
\end{eqnarray}
where $\mathbf{S}_{im}$ is the spin operator of orbital $m$ at site $i$.
Applying DMFT to this model leads to a mapping of the original lattice model
onto an associated quantum impurity problem in a self-consistent bath. In
the particular case of the two-orbital Hubbard model, the associated impurity problem is
the single impurity Anderson model (SIAM) with two levels, where the hybridization function $%
\Delta(\omega)$, which in the usual SIAM is a flat density of states of the
conduction electrons, is now to be determined self-consistently. 

The DMFT iterations consist of the following: Starting from a guessed hybridization  $\Delta(\omega + i\eta)$ for the impurity, its Green function $G(\omega + i\eta)$
is obtained using some numerical method. At this point we introduce the DMRG \cite{garciadmft} to calculate the dynamics using Lanczos \cite{karendin} or the correction vectors for an array of discretized energies $\omega$   \cite{Kunherwhite, Ramasesha}. 
From this we can compute the self energy $\Sigma(\omega + i\eta)=G^{-1}-g_0^{-1}$
where $g_0$ is the non-interacting Green function corresponding to $\Delta(\omega+ i\eta)$.
The self energy allows us to compute the Green function on a lattice, in this case
on a square lattice (SL):
\begin{equation}
G_{SL}(\omega + i\eta)=\sum_{k_x,k_y} \frac{1}{\omega_{sl}(k_x,k_y)+\omega + i\eta -\Sigma(\omega + i\eta)} \mbox{.}
\end{equation}
where $\omega_{sl}(k_x,k_y)= t ( \cos(k_x)+\cos(k_y) )$ with $t=1/2$ to have a band of half-width $D=1$. All energies are given in units of $D$.
The lattice Green function $G_{SL}$ defines a new non-interacting Green function
$g_0^{-1} =G_{SL}^{-1} +\Sigma$
which in turn defines a new hybridization 
$t^2 \Delta(w)= \omega + i\eta  -g_0^{-1}(\omega + i\eta )$
which is the seed to restart the cycle.
The procedure is repeated until converged lattice or impurity Green functions are obtained
(typically between 5 to 10 iterations).  \cite{libroyuriel}

To implement the algorithm we consider \cite{qimiao,mpl} a general representation
of the hybridization function in terms of continued fractions that define a
parametrization of $\Delta(\omega + i\eta)$ in terms of a set of real and
positive coefficients. As we are here dealing with two levels, each one will be hybridized to its own bath.
The hybrization can be written simply as a continued fraction (``chain geometry'' \cite{bullarmp})
\begin{equation}
	\Delta_m(\omega)  = \frac{t^2}{\omega -a_{1m} - \frac{b_{1m}^2}{\omega-a_{2m} - \dots}}
\end{equation}
The $a_{jm}$ and $b_{jm}$ coefficients represent the on-site energy and nearest-neighbor hopping for the sites of a non-interacting chain (not related to the sites of the original lattice, Eqs. (1-3)), with Hamiltonian:
\small
\begin{eqnarray}
	H_{SIAM}&=&H_{at} + \sum_{m,\sigma}\left(\sum_{j=1}^{N_C-1} b_{jm} c^{\dagger}_{jm\sigma} c_{j+1,m\sigma}\right. \nonumber \\&+&
	\left. \sum_{j=1}^{N_C} a_{jm} c^{\dagger}_{jm\sigma} c_{jm\sigma} + t c^\dagger_{1m\sigma} d_{m\sigma}+ h.c.\right)   
\end{eqnarray}\normalsize

Here $H_{at}=\frac{U}{2} N^2 + \frac{J}{2} S^2 -\mu N$, where $N$ and $S$ are the total charge and spin operators of the impurity, $d_{m\sigma}^\dagger$ creates an electron with spin proyection $\sigma$ on orbital $m$ of the impurity and  $c_{im\sigma}^\dagger$ creates an electron at site $i$ of the non-interacting chain of $N_C$ sites. This cut-off on the length of the chains corresponds to a cut-off in the representation of the continued fraction for the impurity and leads to finite-size effects in the spectra.



\begin{figure}
\onefigure[width=8cm]{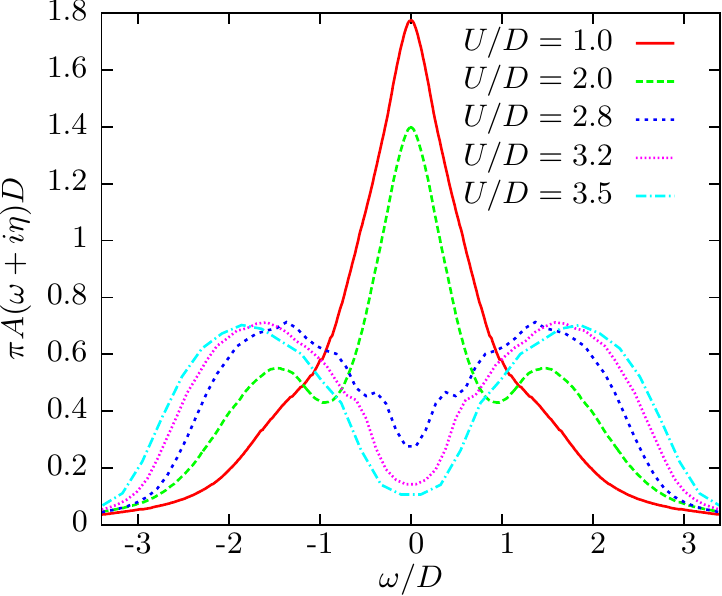}
\caption{(Color online) Converged densities of states
for the half-filled two-band Hubbard model on the square lattice} (1) showing the transition from the metallic to the insulating phases. Here $L=2 N_C+2=42$ sites, $J=-0.1$ and $\eta=0.2$ (which leads to a slightly enhanced DOS in the gap for the insulating states).
\label{fig1}
\end{figure}

In Fig. \ref{fig1}, we show the DMFT+DMRG results for the
densities of states (DOS), $\pi A(\omega + i\eta)=-\operatorname{Im} G(\omega + i\eta)$ for several values of the local interaction $U$ and a finite ferromagnetic $J=-0.1$. We have kept around 256 states in the DMRG decimation procedure and our results were benchmarked with exact diagonalization calculations for smaller systems.

This figure depicts the transition between the insulating (large $U$) and the metallic (small $U$) regimes, showing the lower and upper Hubbard bands in both cases. Since our intention in this paper is to briefly review the new methods developed to solve the impurity within the DMFT, we will not analyze the results in detail, but only mention main physical properties and the comparison between different, albeit complementary, methods.
Here we can clearly see the existence of some structure in the upper and lower Hubbard bands, mainly, the presence of a marked peak in the inner, low-energy, side of the bands. This feature is ubiquitous in recent results for this model and has yet to be fully understood. We can also see an incipient peak in the middle of each band in the insulating regime, which seems to be reminiscent of the van-Hove peak present in the non-interacting half-filled square lattice. In a separate and more detailed paper we will show results for the metal-insulator transition as a function of $J$ \cite{inpreparation}.

\begin{figure}
\onefigure[width=8cm]{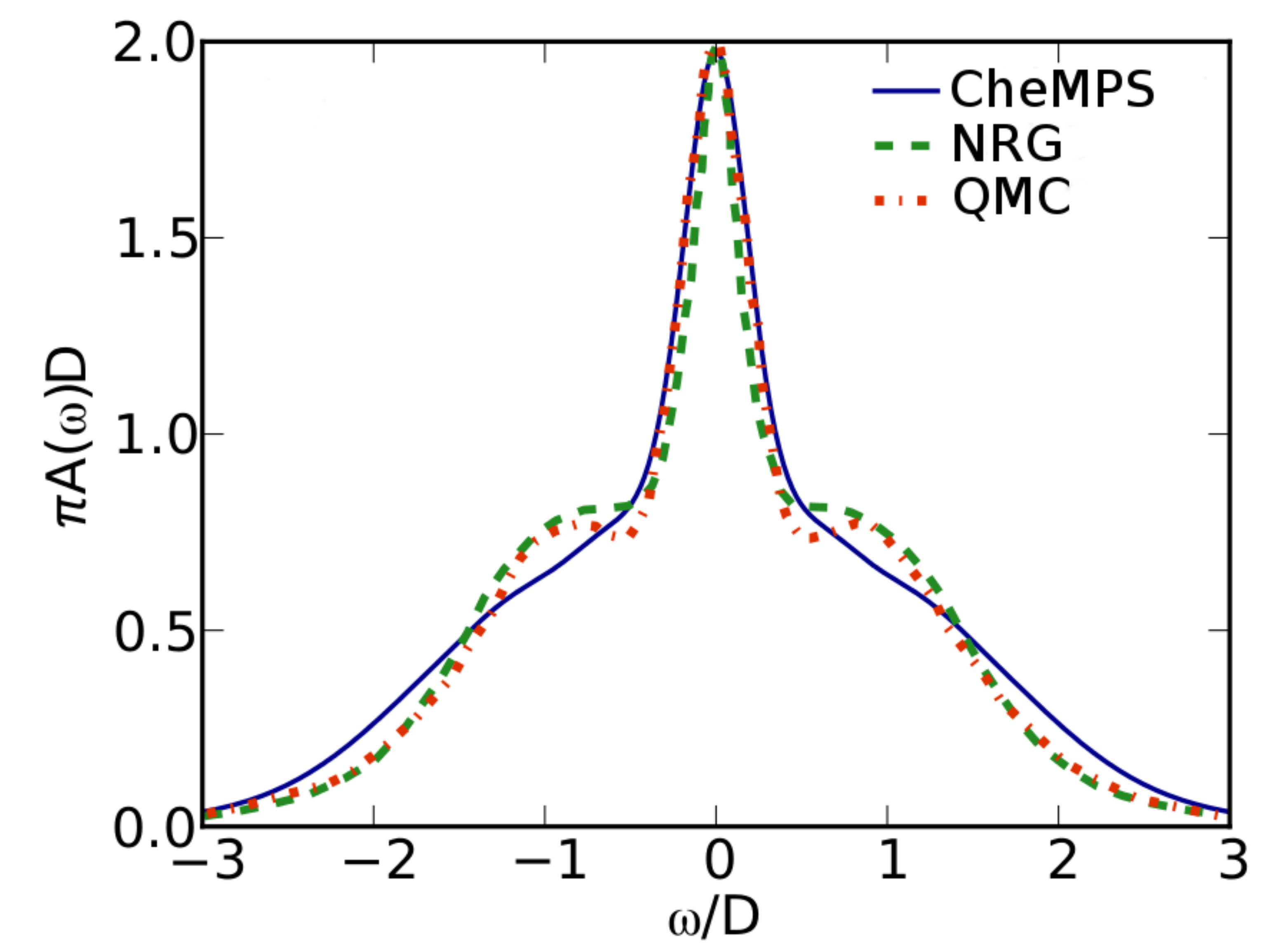}
\caption{(Color online) DMFT results for the spectral density of a half-filled two-band Hubbard model in the metallic regime on the Bethe lattice (extracted from Ref. \cite{wolf} with permission) using three different quantum impurity solvers: Chebyshev-expanded matrix product states, numerical renormalization group, and continuous-time quantum Monte Carlo.}
\label{fig2}
\end{figure}

To illustrate other reliable techniques mentioned in the text, in Fig. \ref{fig2}  we show results using three of the numerical methods mentioned above (Chebyshev expansion of MPS, NRG and analytically continued QMC)  for the calculation of the spectral density for a related two-orbital Hubbard model on the Bethe lattice (see \cite{wolf} and references therein). The  observed disagreement at high energies is due to different broadening convolutions. 

\begin{figure}
\onefigure[width=8cm]{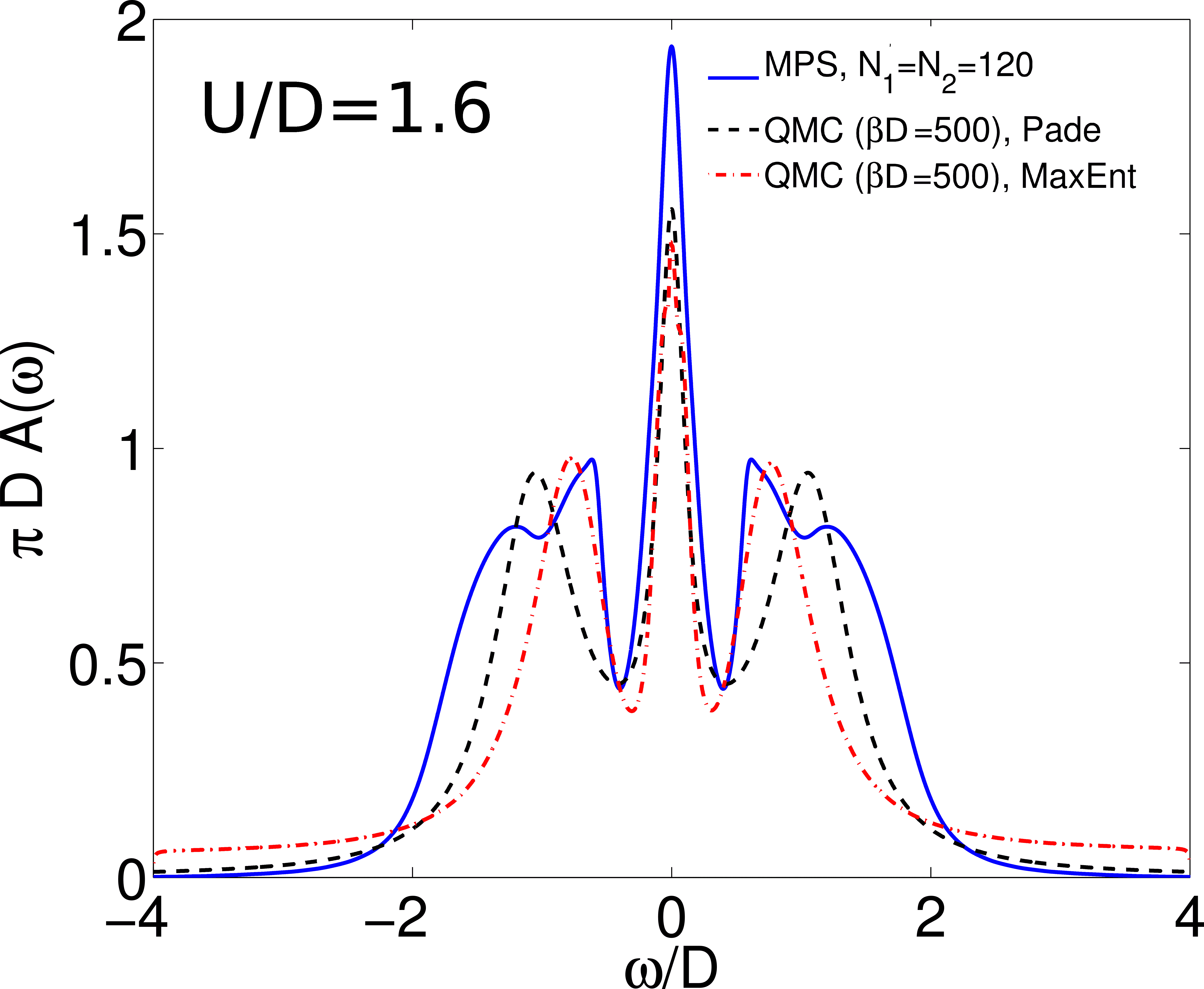}
\caption{(Color online) Comparison of DMFT spectral functions of a half-filled two-orbital Hubbard model in the metallic regime on the Bethe lattice (extracted from Ref. \cite{ganahl} with permission), using other three different quantum impurity solvers: using real-time dynamics together with MPS, QMC+Maximum Entropy and QMC + Pad\'e approximants.}
\label{fig3}
\end{figure}

To complete the comparison between DMFT impurity-solvers based on DMRG-related methods, such as MPS, we show in Fig. \ref{fig3} the spectral functions calculated using other recently developed techniques (for the specific model and parameters used, please refer to \cite{ganahl}). In this case, results based on Fourier transformed real-time dynamics with matrix product states (TEBD) show a much richer structure than previously used methods based on QMC, thus providing a much more reliable tool to study electronic structure in correlated materials. 

\section{Rotationally Invariant Slave-Boson Mean Field Theory }
In the so-called slave boson approach \cite{Coleman1984,KR1986}, the Hilbert space of the on-site Hamiltonian $h_i$ (see Eq. \ref{eq:ham}) is described using an enlarged space which includes fermion (quasiparticle) operators with identical structure as the original (physical) operators, and a set of auxiliary bosonic degrees of freedom. To recover the original Hilbert space, a set of constraints is imposed on the fermionic and bosonic degrees of freedom. The advantages of this approach become clear in the mean field approximation, where the bosonic degrees of freedom are replaced by their mean values. Since the local interaction terms in the Hamiltonian can be represented by a quadratic form in the bosons, the mean field approximation reduces the fermionic action to a quadratic form in the quasiparticle operators, together with a set of parameters (the mean value of the bosonic operators) that can be calculated in a variational way. 
The result is a set of non-interacting fermions having a reduced intersite hopping matrix element due to the interactions. The usual interpretation is that the mean field approach produces a description of the low energy quasiparticles which have a renormalized mass. 

It was shown by Li et al. \cite{li1989spin} studying the single orbital Hubbard model that, in order to correctly describe spin fluctuations,
it is necessary to properly symmetrize the bosonic Hilbert space which in turn makes the theory spin-rotational invariant.
Recently, these ideas have been generalized in a consistent and general way in Ref. \cite{Lechermann2007} where it was also shown that other advantages,
such as the ability to study general forms of the interacting Hamiltonian, or to describe systems in which the quasiparticle weight is non diagonal,
are in close relation with having a rotationally invariant formalism.

In this approach the auxiliary boson operators have two indices $\{\phi_{AB}\}$ associated to eigenstates $|A\rangle$ and $|B\rangle$ of $h_i$ having the same charge parity. 
To obtain a one-to-one mapping with the original Hilbert space, time-independent Lagrange multipliers $\lambda_0$ and $\Lambda$ are used
to enforce the following constraints:
\begin{eqnarray} \label{eq:constr0}
\sum_{A,B,C}\phi_{CA}^\dagger\phi_{CB}\langle B |\hat{O}|A\rangle&=&\hat{O},
\end{eqnarray}
where $\hat{O}=\{1,\, f^\dagger_\al f_\be\}$ and $f^\dagger_\al$ creates a quasiparticle in orbital $\alpha\equiv m\sigma$.

The physical operators are represented by a linear combination of the quasiparticle operators.
\beq \label{eq:physop}
\mathbf{d}^\dagger \to  \mathbf{f}^\dagger R^\dagger
\eeq
where $\mathbf{d}^\dagger=\begin{pmatrix} d_1^\dagger,\ldots,d_M^\dagger\end{pmatrix}$, and the $R$  matrix is a function of the boson fields such that the matrix elements in the new representation are identical to the original ones \cite{Lechermann2007}. 

In the quantum impurity problem, the kinetic energy is accounted for by the coupling of the atomic states to the electron bath which
is described by the hybridization matrix $\nu_d$, and can be written as $H_v=\frac{1}{2} \mathbf{d}^\dagger {\nu}_d \mathbf{d}$.
Using Eq. (\ref{eq:physop}) this can be written in terms of the quasiparticle operators $\mathbf{f}$. We get $H_v =\frac{1}{2} \mathbf{f}^\dagger \nu_f\,  \mathbf{f}$.
To construct the action of the system,  we write the interaction terms as a function of the bosons, and the fermionic action, which is quadratic on the fermion fields, reads:
\begin{eqnarray} 
S_f &=& -\frac{T}{2}\sum_n\mathbf{f}^\dagger(\iom)G^{-1}(\iom)  \mathbf{f}(\iom).
\end{eqnarray}
where $T$ is the temperature,
\beq
G(\iom) = \left[\iom {I}- {\nu_f}\, -{\Lambda}\right]^{-1},
\eeq
and $I$ the identity matrix.
We replace the bosons with time-independent complex numbers and integrate the fermions,
\beq
Z = \text{Tr}\, \left[e^{-S_f}\right]
\eeq
The Free energy $\Omega = -T \ln[Z]$ in the saddle point approximation, including the constraints, reads
\begin{eqnarray}
    \Omega &=&  -T \text{Tr}\ln[-G^{-1}(\iom)] -\lambda_0 + \sum_{ACD}\phi_{AD}^*\left\{\delta_{CD} \lambda_0 \right.\nonumber \\&+ &\left.\delta_{CD} E_A - \langle C| \mathbf{f}^\dagger {\Lambda}\, \mathbf{f}|D\rangle \right\} \phi_{AC}
\label{eq:freenp}
\end{eqnarray}
where $E_A$ is the eigenenergy of state $|A\rangle$.

The saddle-point equations are obtained performing partial derivatives of $\Omega$ with respect to the bosons and the Lagrange multipliers.

\begin{figure}
\onefigure[width=8cm]{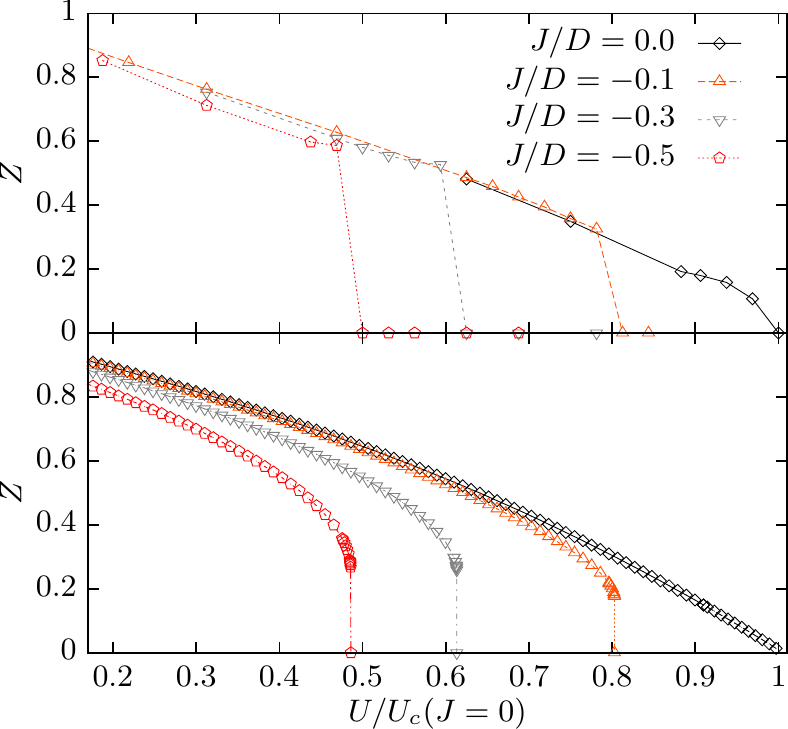}
\caption{(Color online) Quasiparticle weights $Z$ vs $U/U_c(J=0)$. Top: DMRG results, $U_c(J=0)/D = 3.2$.
 Bottom: RISB results for $D/T = 10000$, $U_c(J=0)/D = 4.8$.}
\label{comparacion}
\end{figure}
We have solved the model of Eq. \ref{eq:ham} using RISB as an impurity solver. In Fig. \ref{comparacion}, we compare the results for the quasiparticle weight, $Z$, obtained from the DMRG calculations and the RISB formalism.
It is well known that slave bosons methods tend to overestimate the metalicity of a system. This is reflected in a larger value of $Z$ for a given value of the interaction $U$, and of the critical interaction, $U_c$, at which the metal-insultar transition occurs. The figure shows that in the present problem when $U$ is renormalized by the critical interaction of the $J=0$ calculation, a good agreement between the methods is achieved for non-zero values of $J$. Interestingly, for $J\neq0$, there is a jump from finite values of Z to zero at the metal-insulator transition. We used the RISB formalism to analyze the behavior of the lattice free energy at the transition and confirm that the transition is of first order for finite values of $J$ while it is of second order for $J=0$.

\section{Conclusions and perspectives}
Recent advances in numerical techniques for the resolution of multiorbital quantum impurity problems have made possible the implementation of the Dynamical Mean Field Theory (DMFT) in model Hamiltonians and  Density-Functional-based methods for the calculation of strongly correlated materials in a realistic way.  
Here we review several techniques which have proven useful to calculate spectral densities and other physical properties of these materials.

In particular, we focus on the Density Matrix Renormalization Group (DMRG)-based techniques and compare results with the Rotationally Invariant Slave Boson (RISB) spectra. Both methods are complementary: RISB allows for approximate calculations in the thermodynamic limit of the low energy properties at a relatively low computational cost. DMRG (or its equivalent MPS implementation) is a controlled numerical method which allows for the calculation of spectral densities directly on the real axis at all energy scales. As an example, we solved a two-orbital Hubbard model including a Hund's coupling term $J$ on a square lattice. We found a metal-to-insulator transition as a function of the local repulsion $U$ which is second order for $J=0$ and becomes first order for $J\neq 0$. These new proposals for calculating spectral functions on the real axis show a richer structure in the correlated bands when compared to more traditional methods for respponse functions on the imaginary axis. 

New techniques relying on the optimization of Hilbert spaces from the quantum information perspective \cite{schollmps,Orus} such as the DMRG and extended methods including MPS and tensor networks in general, are being developed. They will be able to tackle more complex multiorbital Hamiltonians and cluster DMFT approximations to treat spatially extended correlations. They are likely to produce more reliable calculations of higher energy spectra with important implications on physical magnitudes such as optical conductivity, and non-equilibrium properties of complex correlated systems. 
 
\subsection{Acknowledgments}
We acknowledge support from PICT2012-1069 of ANPCyT and PIP 0832. This work was partially supported by a grant from the Simons Foundation and the hospitality of the Aspen Center for Physics. We thank G. Kotliar, M. Rozenberg and U. Schollw\"ock for fruitful discussions.


\end{document}